\documentclass[aps,letterpaper,nofootinbib,preprint,showpacs,amsmath,amssymb,pdftex11pt]{revtex4}%

\ifnum\pdfoutput>0
	\usepackage[pdftex]{graphicx}
	\usepackage{epstopdf}
\else
	\usepackage{graphicx}%
\fi
\usepackage{latexsym}%
\usepackage[hypertex]{hyperref}%

\def\cF{{\mathcal F}}
\def\cH{{\mathcal H}}

\def\cM{{\mathcal M}}
\def\cN{{\mathcal N}}
\def\cO{{\mathcal O}}

\def\cW{{\mathcal W}}

\def\Li{{\text{Li}}}

\newcommand{\beq}{\begin{equation}}
\newcommand{\beqn}{\begin{equation}\nonumber}
\newcommand{\eeq}{\end{equation}}
\newcommand{\bea}{\begin{eqnarray}}
\newcommand{\bean}{\begin{eqnarray}\nonumber}
\newcommand{\eea}{\end{eqnarray}}

\begin{document}

\begin{center}
{\bf{\Large Bose Condensation and the BTZ Black Hole}}
\bigskip
\bigskip

{{Cenalo Vaz$^{a,b,}$\footnote{e-mail address: Cenalo.Vaz@UC.Edu},
L.C.R. Wijewardhana$^{b,}$\footnote{e-mail address: Rohana.Wijewardhana@UC.Edu}}}
\bigskip

{\it$^a$RWC and $^b$Department of Physics,}\\
{\it University of Cincinnati,}\\
{\it Cincinnati, Ohio 45221-0011, USA}
\end{center}
\bigskip
\bigskip
\medskip

\centerline{ABSTRACT}
\bigskip\bigskip

Although all popular approaches to quantum gravity are able to recover the Bekenstein-Hawking 
entropy-area law in the thermodynamic limit, there are significant differences in their 
descriptions of the microstates and in the application of statistics. Therefore they can have 
significantly different phenomenological implications. For example, requiring indistinguishability 
of the elementary degrees of freedom should lead to changes in the black hole's radiative 
porperties away from the thermodynamic limit and at low temperatures. We demonstrate this 
for the Ba\~nados-Teitelboim-Zanelli (BTZ) black hole. The energy eigenstates and statistical 
entropy in the thermodynamic limit of the BTZ black hole were obtained earlier by us via 
symmetry reduced canonical quantum gravity. In that model the BTZ black hole behaves as a system 
of Bosonic mass shells moving in a one dimensional harmonic trap. Bose condensation does not 
occur in the thermodynamic limit but this system possesses a finite critical temperature, 
$T_c$, and exhibits a large condensate fraction below $T_c$ when the number of shells is finite. 
\bigskip

\vfill\eject

\section{Introduction}
A solution to the problem of explaining black hole thermodynamics \cite{bek72,bch73,haw75} 
from a canonical or microcanonical ensemble can be expected to provide important insights into 
quantum gravity, which is considered to be a central problem of theoretical physics. However, 
there is no general consensus on the true nature of the elementary degrees of freedom 
because many apparently different approaches to quantum gravity have all successfully 
recovered the Bekenstein-Hawking entropy.
 
The differences between the approaches appear fundamental and range from their intrinsic 
description of the quantum gravity states to the statistics used in counting them. For 
example, in string theory the microstates are dual to weak field D-brane states 
\cite{stva96,hrpl97,dab05}, in the AdS/CFT approach they are taken to be the states of 
a particular horizon conformal field theory (CFT) \cite{hkl99,emp99,hmst01,mupa02,chgs07} 
and in loop quantum gravity (LQG) they are represented by punctures of a spin network 
on the event horizon \cite{rov96,abck97,dle04,mei04,cor07,bar08}. It seems unlikely that these 
descriptions are related in some as yet unknown way because of a more subtle distinction 
between them, which is their use of statistics. In string theory and AdS/CFT the elementary 
degrees of freedom are indistinguishable whereas they must be treated as distinguishable in 
LQG in order to recover the Bekenstein-Hawking law. The statistical properties of the microstates 
can have far ranging phenomenological implications \cite{kie07}, particularly at low 
temperatures and outside the thermodynamic limit. 

In this paper we ask what behavior one may expect outside the thermodynamic limit. Such a 
question can only be answered in a model dependent way, therefore we pose it within 
the context of a model of the quantum black hole that arises from canonical quantum gravity
applied to spherical gravitational collapse \cite{vaz01}. In this model a black hole microstate 
is viewed as a particular distribution of matter shells, which have crossed the horizon, among 
the available black hole energy levels. This is a natural way to realize Bekenstein's original 
ideas in \cite{bek72}, from which we quote: ``It is then natural to introduce the concept of 
black hole entropy as the measure of the {\it inaccessibility} of information (to an exterior 
observer) as to which particular internal configuration of the black hole is actually realized 
in a given case.'' 

The simplest example is provided by the static Ba\~nados-Teitelboim-Zanelli (BTZ) black hole
\cite{btz94}. A symmetry reduced canonical quantization of this black hole yields Hawking radiation 
\cite{vaz07} and an equispaced black hole mass spectrum. Furthermore, counting the black hole'
s microstates in the canonical ensemble reproduces the Bekenstein-Hawking entropy in the thermodynamic 
limit, provided that the counting proceeds using Bose statistics \cite{vaz08}. The thermodynamic 
limit is the limit in which both the system size and energy are simultaneously taken to be very 
large. For the BTZ black hole this means that the temperature is high. Our interest here is to 
examine the same ensemble at low temperatures and for a finite number of matter shells. We will 
show that Bose-Einstein condensation occurs and the BTZ black hole turns ``cold'' once condensation 
sets in. ``Cold'' black holes are of course stable against Hawking evaporation. 

One may criticize our choice of the 2+1 dimensional black hole, even though it has the virtue 
of being simple, because it is topological and therefore unique in many respects. For instance when canonical 
quantum gravity is applied to the Schwarzschild or higher dimensional black holes in the absence of a 
cosmological constant the microstates are counted in an ``area ensemble'' using Boltzmann statistics. 
However the situation is different for ``large'' Anti-de Sitter (AdS) black holes, which are black 
holes whose horizon radius is much larger than the AdS length. In a particular limit, their mass spectra 
are identical to the spectrum of the BTZ black hole modulo dimension dependent constant factors and 
Bose statistics must be applied to recover the Bekenstein-Hawking area law \cite{vaz09}. Moreover, 
many higher dimensional black holes from string theory have the form BTZ $\times~ \cM$, where $\cM$ is 
a simple manifold. Their thermodynamic properties can be recovered directly from the thermodynamic 
properties of the BTZ black hole \cite{hyu98,sfsk98}. Bose condensation is a possibility whenever 
indistinguishability is an essential feature of the counting of microstates. It may thus occur for 
realistic black holes in higher dimensions, in which case one can expect stable black hole remnants 
to contribute significantly to the Dark Matter content of the universe. 

This paper is organized as follows. In section II we review the canonical quantization of the 
BTZ black hole and its statistical thermodynamics in the canonical ensemble, as given in \cite{vaz07}. 
We then go outside the thermodynamic limit in section III, treating the black hole as a system of 
finite size. With the black hole spectrum, which we derive in section II, we show that Bose 
condensation can occur. We go on to determine the critical temperature and the condensate fraction 
as well as the entropy and argue that the area law strictly holds only in the thermodynamic limit 
but breaks down outside it and below the critical temperature. We conclude in section IV with some 
comments.

\section{Statistical Thermodynamics of the BTZ Black Hole}

The static BTZ black hole may be viewed as the end state of the collapse of inhomogeneous dust 
\cite{man93,gut05}, which is described by the LeMa\^\i tre-Tolman-Bondi family of solutions of 
Einstein's equations with a negative cosmological constant. The solutions are characterized by 
two arbitrary functions, {\it viz.,} the ``mass function'', $F(\rho)=4GM(\rho)$,  representing 
the initial mass distribution of the dust ball and the energy function, $E(\rho)$ representing 
its initial energy distribution. The classical solutions are given by the circularly symmetric 
line element
\beq
ds^2 = d\tau^2 -\frac{(\partial_\rho R)^2}{2(E-F)}d\rho^2 - R^2 d\varphi^2,
\label{btz}
\eeq
where $\tau$ is the dust proper time and $\rho$ labels dust shells or curvature radius $R(\tau,\rho)$.
Einstein's equations can be integrated to give the energy density of the dust
\beq
\varepsilon(\tau,\rho) = \frac{\partial_\rho F}{R(\partial_\rho R)}
\label{energy}
\eeq
and a dynamical equation for the evolution of the area radius,
\beq
(\partial_\tau R)^2=2E-\Lambda R^2
\label{ein2}
\eeq
in terms of two arbitrary integration functions of the shell index coordinate, $\rho$. The last 
determines the area radius of shells to be
\beq
R(\tau,\rho)=\sqrt{\frac{2E}{\Lambda}}\sin\left(-\sqrt{\Lambda}\tau+\sin^{-1}\sqrt{\frac\Lambda{2E}}
\rho\right),
\label{Req}
\eeq
where we have used a freedom in the scaling of the shell index $\rho$ by setting $R(0,\rho)=\rho$
and the fact that collapse solutions satisfy $\partial_\tau R <0$. The integration function $F(\rho)$ 
represents the initial mass distribution of the collapsing dust ball and the function $E(\rho)$ 
represents the initial energy distribution; one finds
\bea
&&F(\rho)=\int_0^\rho \varepsilon(0,\rho)\rho d\rho\cr\cr
&&E(r) = \frac 12 \left[(\partial_\tau R)^2_{\tau=0} + \Lambda R^2\right].
\eea
directly from \eqref{energy} and \eqref{ein2} respectively.

The general circularly symmetric Arnowitt-Deser-Misner (ADM) line element,
\beq
ds^2 = N^2 dt^2 - L^2 (dr-N^r dt)^2 -R^2 d\varphi^2
\eeq
can be embedded into the metric in \eqref{btz}. After a series of transformations described in detail in
\cite{vaz07} this leads to a canonical description of the classical black hole in terms of the 
dust proper time, $\tau(r)$, the area radius, $R(r)$, the mass density defined via
\beq
F(r) = \frac{M_0 }2 + \int_0^r dr'~ \Gamma(r),
\eeq
where $M_0$ is an arbitrary constant contributed by the boundary at the origin, and their 
conjugate momenta, $P_\tau(r)$, $P_R(r)$ and $P_\Gamma(r)$ respectively. The effective constraints of 
the gravity-dust system are then obtained in the form
\bea
\cH_r &=& \tau' P_\tau + R' P_R - \Gamma P_\Gamma' \approx 0\cr\cr
\cH &=& P_\tau^2 + \cF P_R^2 -\frac{\Gamma^2}{\cF} \approx 0,
\label{newconst}
\eea
where $\cF = \Lambda R^2 -F$ and the prime refers to derivatives with respect to the ADM label 
coordninate $r$. 

Applying Dirac's formal quantization to the above system of constraints, we replace the momenta 
by functional derivatives with respect to their corresponding configuration variables. To 
encapsulate the factor ordering ambiguities at the formal level of the Wheeler-DeWitt equation 
we introduce factors of $\delta(0)$ into the resulting functional Schroedinger equation, writing 
the quantum Hamiltonian constraint 
as 
\beq
{\widehat \cH} \Psi[\tau,R,\Gamma] = \left[\frac{\delta^2}{\delta \tau^2} + 
\cF \frac{\delta^2}{\delta R^2} + A\delta(0) \frac{\delta}{\delta R} + B\delta(0)^2+
\frac{\Gamma^2}{\cF}\right] \Psi[\tau,R,\Gamma] = 0,
\label{qham}
\eeq
where $A(R,F)$ and $B(R,F)$ are smooth functions. The continuum limit of the wave-functional is 
taken to be of the form
\beq
\Psi[\tau,R,\Gamma] = \exp\left[i\int dr \Gamma(r) \cW(\tau(r),R(r),F(r))\right],
\label{wfnal}
\eeq
which formally obeys the momentum constraint provided that $\cW(\tau,R,F)$ has no explicit dependence 
on the label coordinate $r$. We then put \eqref{qham} and \eqref{wfnal} on a lattice and take the 
continuum limit \cite{vaz04}. This corresponds to the choice of regularization. Consistency 
of the lattice regularization fixes the factor ordering via a set of three equations \cite{vaz06} 
for each lattice site, $j$,
\bea
&&\left[\left(\frac{\partial \cW_j}{\partial\tau_j}\right)^2 + \cF_j 
\left(\frac{\partial \cW_j}{\partial R_j}\right)^2 - \frac{1}{\cF_j}\right]=0,\cr\cr
&&\left[\frac{\partial^2 \cW_j}{\partial\tau_j^2} + \cF_j\frac{\partial^2 \cW_j}
{\partial R_j^2} + A_j \frac{\partial \cW_j}{\partial R_j} \right]  =  0,\cr\cr
&& B_j = 0.
\label{4eqns}
\eea
Thus $B(R,F)$ is constrained to be identically vanishing and $A(R,F)$ and the lattice wave-functions 
are obtained by solving the first two equations. Furthermore, hermiticity of the Hamiltonian constraint 
requires that the Hilbert space measure $\frak{m}(R,F)$ is determined from $A(R,F)$ according to
\beq
A_j = |\cF_j|\partial_{R_j} \ln (\frak{m_j}|\cF_j|).
\label{meas}
\eeq
Thus, the system is completely solved once regularized on the lattice.

Unfortunately, the situation becomes more complicated when the mass density function is distributional, 
as it is in the description of the final state black hole, for then the system naturally collapses into 
a countable product of wave functions, the functional differential equations become ordinary partial
differential equations and no regularization is required. It follows that no further conditions are 
available, leaving the factor ordering ambiguity in the form of the unknown functions $A$ and $B$ 
unresolved. 

This is exemplified by the BTZ black hole whose metric is of the form
\beq
ds^2 = - \left(\Lambda R^2-8GM\right) dT^2 + \frac{dR^2}{(\Lambda R^2 -8GM)} + R^2 d\varphi^2,
\label{btzmetric}
\eeq
where $M$ is the black hole mass parameter. This is a special solution of \eqref{btz} in which the mass 
function is taken to be constant, $F=4GM$, for $\rho>0$ and the energy function is given by $2E=1+8GM$, 
again for $\rho>0$. The metric in \eqref{btz} can be brought to the static form in \eqref{btzmetric}
by the transformations $R=R(\tau,\rho)$ as given in \eqref{Req} and 
\beq
T = \tau + \int dR \frac{\sqrt{1+8GM-\Lambda R^2}}{\Lambda R^2-8GM}
\eeq
for the Killing time, $T$. Within the framework of the canonical theory we take
\beq
F(r) = 4GM_0+4G\varepsilon \Theta(r) = 4GM,
\label{massfn2}
\eeq
where $\varepsilon$ represents the mass of a shell at $r=0$ and where $\Theta$ is the Heaviside function. 
Likewise, the energy function should be given by
\beq
E(r) = \frac 12 \left[1+8GM_0+8G\varepsilon\Theta(r)\right].
\label{efn2}
\eeq
We see that the mass function in \eqref{massfn2} yields a mass density that is the $\delta-$distribution 
\beq
\Gamma(r) = 4G\varepsilon\delta(r),
\label{Gammabtz}
\eeq
and the wave-functional in \eqref{wfnal} turns into the wave-{\it function}, 
\beq
\Psi = e^{\frac i{4G}\int_0^\infty dr \Gamma(r) \cW(\tau(r),R(r),F(r))} = e^{i\varepsilon \cW(\tau,R,F)},
\eeq
where $\tau=\tau(0)$, $R=R(0)$ and $F=F(0)$. The Wheeler--DeWitt equation becomes 
\beq
\left[\frac{\partial^2}{\partial \tau^2}+\cF\frac{\partial^2}{\partial R^2} + A
\frac{\partial}{\partial R}+B\right]e^{i\varepsilon\cW(\tau,R,F)} = 0
\label{wd1}
\eeq
and we note that $A(R,F)$ and $B(R,F)$ remain undetermined, as does the measure $\frak{m}(R,F)$, 
with $B(R,F)$ playing the role of an external potential.

Fortunately there is a way out of this difficulty, which is provided by a detailed examination of the 
Hawking radiation from the black hole within the context of this midisuperspace model. 
Because the Wheeler DeWitt equation is second order in the dust proper time we can define both positive 
and negative frequency states with respect to $\tau$. Therefore, taking the collapsing dust as a small 
perturbation around a pre-existing, massive black hole, so that the role of the quantum matter in 
Hawking's original background field derivation of black hole radiance \cite{haw75} is played by the 
dust, the system \eqref{4eqns} can be shown to yield Hawking radiation at the Hawking temperature 
\cite{vaz07}. However, an important point is that obtaining  the correct Planckian distribution in 
the near horizon limit requires a particular choice of measure, one appropriate to the massive black hole
because the dust itself is considered a perturbation on the black hole background. The measure turns out to 
be the one obtained from the deWitt supermetric, which, from \eqref{newconst}, is
\beq
\gamma_{ab} = \left(\begin{matrix}
1 & 0\cr
0 & \frac 1\cF
\end{matrix}\right).
\eeq
Thus 
\beq
\frak{m}=\frac 1{\sqrt{|\cF|}}
\eeq
and putting this together with the hermiticity requirement in \eqref{meas} determines $A(R,F)$. Now
since the black hole ends up being a single shell in this simple quantum mechanical model, it is 
reasonable to take the external potential, $B$, to be vanishing. Thus \eqref{wd1} reduces to the free 
Klein-Gordon equation which is hyperbolic in the interior and elliptic elsewhere. A stationary state 
solution this equation was shown in \cite{vaz08} to yield the mass levels
\beq
\varepsilon_j = \frac{\hbar}l \left(j+\frac 12\right),
\label{spectrum}
\eeq
where $l^2=-\Lambda^{-1}$ is the AdS length. 

The macroscopic black hole is defined as the end state of the collapse of many, say $\cN=\sum_j\cN_j$, 
matter shells, $\cN_j$ of which occupy level $j$ of \eqref{spectrum}. Thus, when a boundary contribution 
from the origin  is included \cite{man93,gut05}\footnote{In more than three dimensions this contribution 
from the origin is usually set to zero, otherwise it would represent a singular initial configuration. 
In three dimensions a non-vanishing $M_0$ is necessary to allow for a velocity profile that vanishes at 
the origin. This does not lead to singular initial data and the presence of $M_0$ does not lead 
to a singular initial configuration.}, the total black hole mass is given as 
\beq
M = M_0 + \sum_j \frac{\hbar}l \left(j+\frac 12\right) \cN_j.
\eeq
From this point of view, the elementary degrees of freedom are bosonic mass shells and a 
black hole microstate is a particular distribution of $\cN$ shells between the levels in 
\eqref{spectrum}.

The statistical thermodynamics, specifically the $\cN\rightarrow \infty$ limit, of this model 
can be treated in the canonical ensemble and is captured by the partition function
\beq
Z(\beta) = \sum_{\{\cN_1,\ldots,\cN_j,\ldots\}}g(\cN_1,\ldots,\cN_j,\ldots) \exp
\left[-\beta\left(M_0+ \sum_{j}\varepsilon_j \cN_j\right)\right],
\eeq
where $\cN_j$ represents the number of shells excited to level $j$, with mass $\varepsilon_j$, 
and $g(\cN_1,\ldots,\cN_j,\ldots)$ is the degeneracy of states, which we take to be unity
so as to implement Bose statistics. The canonical entropy is then obtained from
\beq
\left. S_\text{can} = [\beta M + \ln Z(\beta)\right]_{M=-\partial \ln Z/\partial \beta},
\label{ent1}
\eeq
where $M$ is the average energy in the canonical ensemble, which we associate with the 
black hole mass. When the system size is taken to be infinite, the partition function
\beq
Z(\beta\hbar/2l) = e^{-\beta
  M_0}\prod_{j=0}^\infty\left[1-e^{-\frac{\beta\hbar}{2l}
    \left(2j+1\right)} 
\right]^{-1}.
\eeq
is easily evaluated. It may be rewritten, by exploiting the well known duality in \cite{hr18}, as
\beq
Z(\beta\hbar/2l) = \frac 1{\sqrt{2}}e^{-\left(\frac{\pi^2
      l}{\beta\hbar}+\frac{\beta\hbar}{8l}\right)\left( 
\frac{8lM_0}{\hbar}-\frac 16\right)} [Z(4\pi^2 l/\beta\hbar)]^{-1}.
\label{modular}
\eeq
This links the high temperature behavior of our system to its low temperature dynamics. If 
one assigns a value $\Delta_0$ to the ground state energy of the system, taking 
\beq
Z(4\pi^2 l/\beta\hbar) \approx e^{-\frac{8\pi^2 l^2\Delta_0}{\beta\hbar^2}}
\eeq
then it is then a straightforward exercise to show that
\beq
S_\text{can}= 4\pi \sqrt{c_\text{eff} \frac{lM}{6\hbar}}
\label{Scan}
\eeq
where
\beq
c_\text{eff} = \frac 12 \left[1-\frac{48l}{\hbar}(M_0-\Delta_0)\right]
\eeq
depends on two undetermined parameters, the vacuum energy $\Delta_0$ and the boundary contribution
$M_0$. The vacuum 
energy depends on what one takes to be the ground state of the BTZ solution, for example the 
choice $\Delta_0=-1/8G$ corresponds to the choice of pure AdS$_3$, 
\beq
ds^2= -\left(\frac{R^2}{l^2}+1\right)dT^2 + \left(\frac{R^2}{l^2}+1\right)^{-1}
dR^2 + R^2d\varphi^2,
\label{M02}
\eeq
for the ground state \cite{bhtz93}. On the other hand, the presence of the arbitrary contribution, 
$M_0$, from the boundary at the origin must be fixed by some other means, for example by comparing 
\eqref{Scan} to the Bekenstein-Hawking entropy of the BTZ black hole. The Bekenstein-Hawking
entropy is just
\beq
S_\text{B-H} = \frac{A_h}{4G} = \frac{\pi R_h}{2G\hbar} = \frac{\pi l}{\hbar}\sqrt{\frac{2M}G},
\eeq
where $A_h$ is the horizon area and $R_h$ its radius. This in turn implies that
\beq
c_\text{eff} = \frac{3l}{2G\hbar},
\eeq
and therefore fixes $M_0-\Delta_0$.

\section{The Grand Canonical Ensemble}

We now consider a black hole made of a finite number $\cN=\sum_j\cN_j$ shells. It is standard 
practice to then begin with the grand partition function,
\beq
\Xi(\varepsilon,\beta) = e^{-\beta M_0} \prod_j (1-e^{-\beta(\varepsilon_\alpha-\mu)})^{-1},
\eeq
where we introduced a chemical potential $\mu$, which is determined from the constraint 
\beq
\sum_j \langle \cN_j\rangle  = \cN
\eeq 
and  where $\varepsilon_j$ is the energy of the level $j$. Introducing 
$\lambda=e^{\beta(\mu-\varepsilon_0)}<1$ ($\lambda^{-1}$ is the fugacity) and $\Delta
\varepsilon_j=\varepsilon_j-\varepsilon_0$ we then have
\beq
\Xi(\mu,\beta) = e^{-\beta M_0}\prod_j \frac\lambda{\lambda-e^{-\beta\Delta
\varepsilon_j}}.
\eeq
The average level occupation number is computed directly from $\Xi(\beta,\mu)$,
\beq
\langle N_j\rangle = -\frac 1\beta \frac{\partial\ln \Xi}{\partial \varepsilon_j} 
= \frac 1{e^{\beta\Delta\varepsilon_j}-\lambda}
\eeq
so our condition for determining the chemical potential is
\beq
\cN=\sum_{j=0}^\infty \frac 1{e^{\beta\Delta\varepsilon_j}-\lambda} = 
\frac{\lambda}{1-\lambda} + \sum_{j=1}^\infty \frac \lambda{e^{\beta\Delta\varepsilon_j}
-\lambda}.
\label{num}
\eeq
The first term on the right represents the ground state occupancy (henceforth, average 
values will be understood and the angular brackets will be omitted),
\beq
N_0=\frac \lambda{1-\lambda},
\eeq
and the second represents the number of excited shells,
\beq
N_\text{ex} = \sum_{j=1}^\infty \frac \lambda{e^{\beta\Delta\varepsilon_j}-
\lambda}.
\eeq
The condensate fraction is defined as the ratio of the ground state occupancy to the 
total number of shells, $f_C=N_0/\cN$. 

We will first determine $f_C$ as a function of the temperature. For the spectrum of the 
BTZ black hole in \eqref{spectrum}, $\Delta \varepsilon_j=\hbar j/l$ and therefore, replacing 
the last sum in \eqref{num} by its power series expansion, we can express the total number 
of shells as
\beq
\cN = \frac\lambda{1-\lambda} + \sum_{j=1}^\infty\sum_{r=1}^\infty 
\lambda^r e^{-\beta \hbar rj/l}.
\label{num2}
\eeq
The most direct way to evaluate the double sum in \eqref{num2} is to employ the Mellin-Barnes 
representation of the exponential function,
\beq
e^{-\alpha} = \frac 1{2\pi i} \int_{\tau-i\infty}^{\tau+i\infty} dt~ 
\Gamma(t) \alpha^{-t}
\label{mellinbarnes}
\eeq
where $\tau \in \mathbb{R}$ and $\text{Re}(\alpha)>0$, to re express it as an integral 
over the complex plane. This gives
\beq
\cN = \frac\lambda{1-\lambda} + \frac 1{2\pi i} 
\int_{\tau-i\infty}^{\tau+i\infty} dt~ \frac{\Gamma(t)\zeta(t)\Li_t
(\lambda)}{(\beta\hbar/l)^t},
\eeq
where $\zeta(t)$ is the zeta function and $\Li_t(\lambda)$ is the polylogarithm. For 
$\lambda<1$ the integral is governed by the simple pole of the $\zeta-$function at 
$t=1$, therefore 
\beq
\cN = \frac \lambda{1-\lambda} + \frac{\Li_1(\lambda)}{\beta\hbar/l} 
\equiv \frac \lambda{1-\lambda} - \frac 1{\beta\hbar/l}\ln(1-\lambda).
\label{num3}
\eeq
Equation \eqref{num3} can be solved for $\lambda$ in terms of the Lambert function, 
$\cW(x)$, 
\beq
\lambda(\beta,\cN) = 1 - \frac{\beta\hbar/l}{\cW(\beta\hbar/l 
e^{(\cN+1)\beta\hbar/l})}
\label{fugacity}
\eeq
and determines the ground state occupancy as well as the condensate fraction,
\beq
f_C(\beta,\cN) = \frac{N_0}\cN = \frac{\cW(\beta\hbar/l e^{(\cN+1)\beta\hbar/l})}
{\cN\beta\hbar/l}-\frac 1\cN,
\eeq
in terms of the temperature. It is more illuminating, however, to express the condensate 
fraction as a function of the dimensionless ratio, $x=T/T_c$, where $T_c$ is the critical 
temperature at which all the shells are excited.

To calculate the critical temperature, $T_c$, it is necessary to set $\lambda \approx 1$ 
and $N_\text{ex}=\cN$, the total number of shells. Once again employing the Mellin-Barnes 
transformation, this time replacing the polylogarithm function 
$\Li_t(\lambda)\vert_{\lambda\approx 1}$ by the $\zeta-$function, we find \cite{hol98}
\beq
N_\text{ex} = \cN \approx \frac 1{2\pi i} \int_{\tau-i\infty}^{\tau+i\infty} 
dt~ \frac{\Gamma(t)\zeta^2(t)}{(\beta_c\hbar/l)^t},
\eeq
but the integral now admits a double pole at $t=1$ via the $\zeta-$function and therefore 
\beq
\cN \approx \left(\frac{\Gamma(t)}{(\beta\hbar/l)^t}\right)' = - \frac 
l{\beta_c\hbar}\ln (e^\gamma \beta_c \hbar/l),
\eeq
where $\gamma$ is the Euler-Mascheroni constant. Solving for the inverse critical 
temperature, we determine
\beq
\beta_c\hbar/l = \frac{\cW(e^{-\gamma}\cN)}{\cN},
\label{crtemp}
\eeq
which is the exact version of the result in \cite{dal99}. The Lambert function increases 
linearly for small values of its argument and slower than the log function for larger 
values. Thus, for a few shells the critical temperature is more or less constant whereas 
it increases roughly as $\cN/\ln \cN$ when $\cN$ grows large. This is a well known property of 
one dimensional harmonic traps.
\begin{figure}
\begin{minipage}[t]{7cm}
\includegraphics[width=7cm]{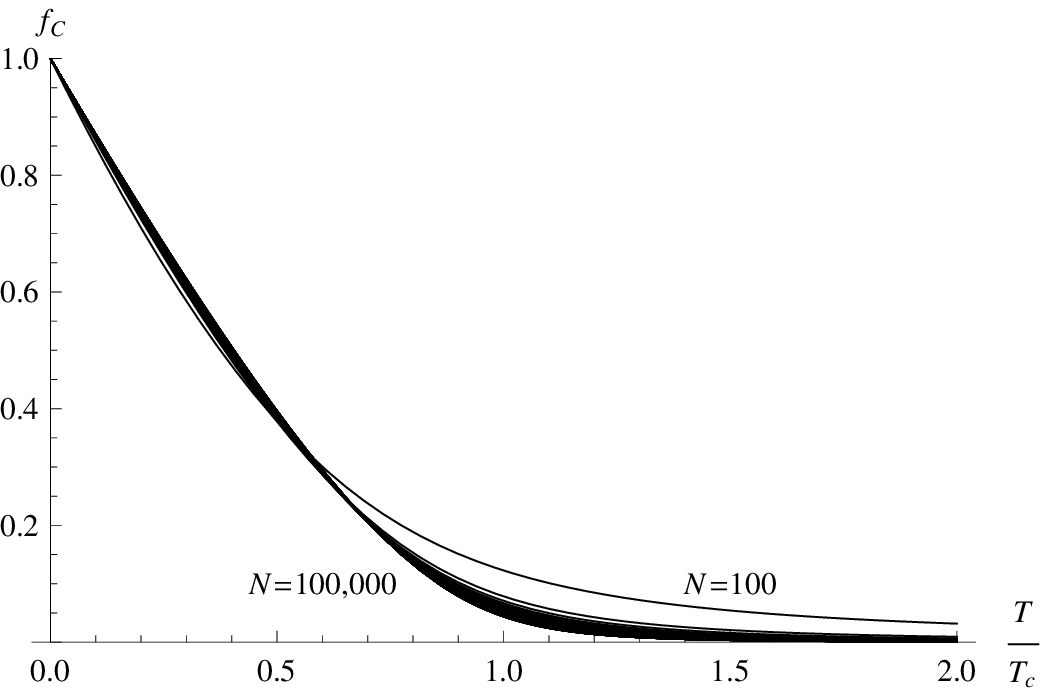}
\caption{$f_C$, as a function of $T/T_c$.}
\label{condensate}
\end{minipage}
\hfil
\begin{minipage}[t]{7cm}
\includegraphics[width=7cm]{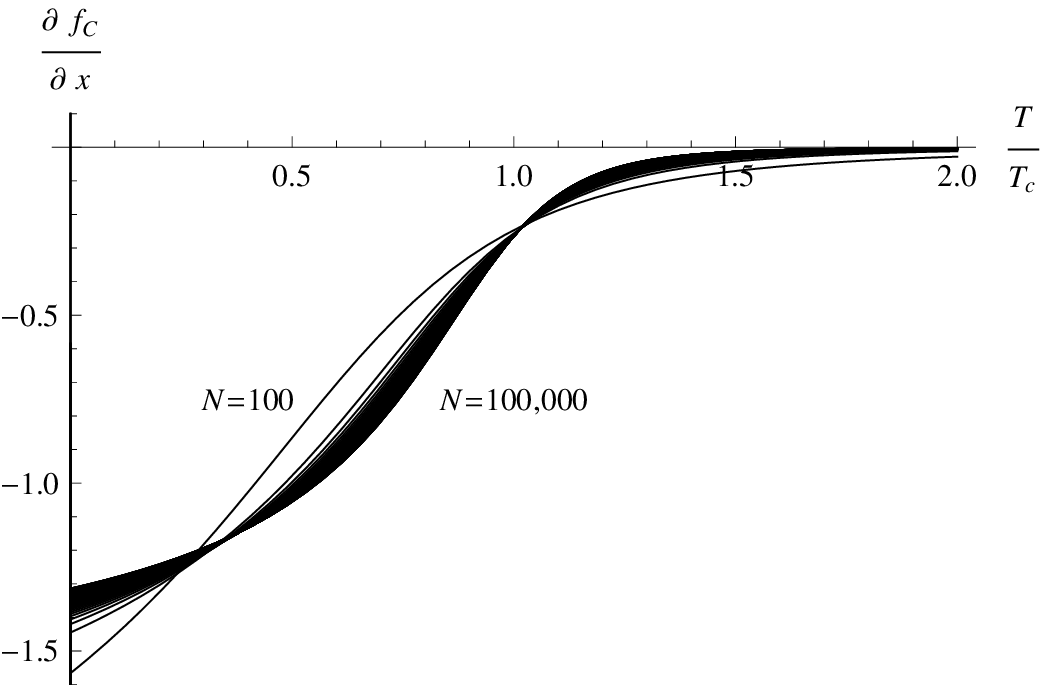}
\caption{$\frac{\partial f_C}{\partial x}$, as a function of $T/T_c$.}
\label{slope}
\end{minipage}
\end{figure}
When the black hole is composed of many shells, most of them will in fact lie in the 
ground state. However, for a small number of shells the ground state occupancy can still be 
a large percentage of the total number of shells in a significant interval of temperatures.

Using \eqref{crtemp} to re-express the condensate fraction in terms of the variable $x=T/T_c$, 
we easily arrive at
\beq
f_C(x,\cN) = \frac{x}{\cW_N} \cW(\frac{\cW_N}{\cN x} e^{\frac{(\cN+1)\cW_N}{\cN x}})-
\frac 1\cN,
\eeq
where we have defined $\cW_N = \cW(e^{-\gamma} \cN)$. At high temperatures,
\beq
f_C \stackrel{x\rightarrow \infty}{\approx} \frac 1\cN+\frac{\cW_N}{\cN x} + \cO(x^{-2})
\eeq
whereas, at low temperatures,
\beq
f_C \stackrel{x\rightarrow 0}{\approx} 1 - \frac{\ln \cN}{\cW_N} x + \cO(x^2)
\eeq
Figure \ref{condensate} shows the behavior of $f_C$ as a function of $x$ for $\cN$ ranging from 
one hundred to one hundred thousand shells. It is small, increasing slowly to the left when 
$x \gg 1$ but significantly faster when $x\ll 1$ and approaching unity in the limit as $x 
\rightarrow 0$. Again, the slope of the condensate fraction, $\partial f_C/\partial x$, approaches 
$\ln \cN/\cW_N$ for small $x$ and it is vanishingly small for large $x$. Figure \ref{slope} shows 
this slope for the same range of $\cN$. Both the 
analytical approximations above as well as the figures indicate that increasing the number of 
shells saturates the slope of the condensate fraction. This behavior suggests that the critical 
temperature may be thought of as the condensation temperature. 

The average energy of the system, which is associated with the black hole mass, can be 
expressed in terms of the occupancy of the levels in the usual way and the sum rewritten 
in the integral representation we have used so far; one gets
\beq
M(\beta,\cN) = M_0 + \cN\varepsilon_0 + \frac{\hbar/l}{2\pi i} \int_{\tau-i\infty}^{\tau+
i\infty} dt~ \frac{\Gamma(t)\Li_t(\lambda)\zeta(t-1)}{(\beta\hbar/l)^t}.
\eeq
For $\lambda<1$ there is only a simple pole coming from the $\zeta-$function 
at $t=2$, so the black hole mass can be given in terms of the temperature 
by 
\beq
M(\beta,\cN) = M_0 + \cN \varepsilon_0 + \frac{\Li_2(\lambda)}{\beta^2\hbar/l}
\label{mass}
\eeq
and the heat capacity is
\beq
C(\beta,\cN) = \frac{2\Li_2(\lambda)}{\beta\hbar/l} - \frac{\Li_2'(\lambda)}{\hbar/l} 
\left(\frac{\partial\lambda}{\partial\beta}\right)_\cN,
\eeq
where the prime on the polylogarithm represents a derivative with respect to $\lambda$. 
\begin{figure}
\includegraphics[width=7cm]{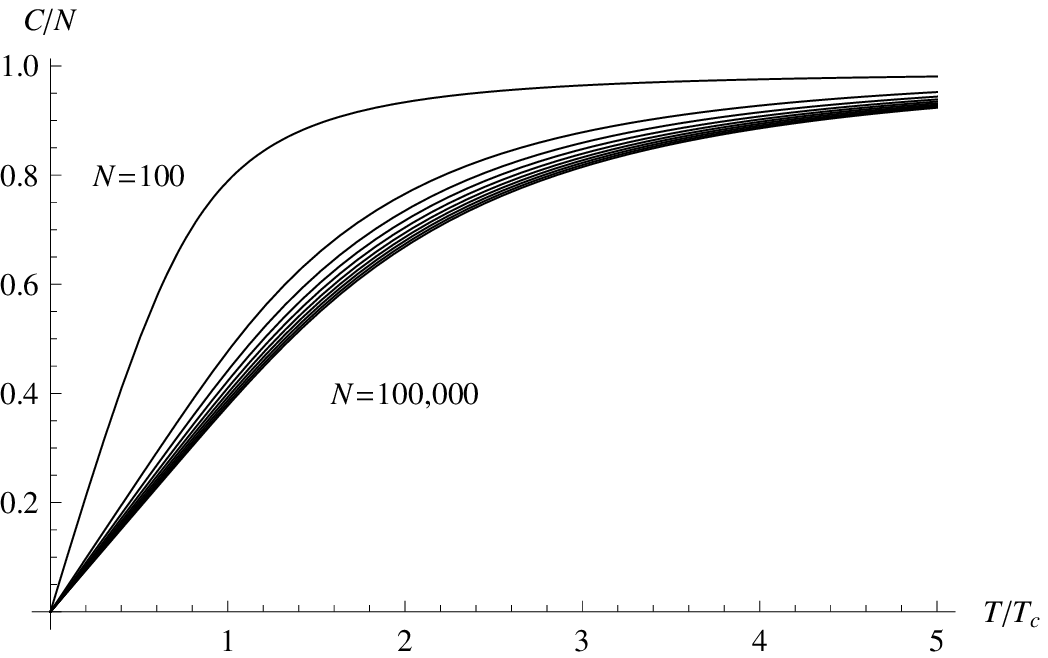}
\caption{$C/\cN$ as a function of $T/T_c$.}
\label{heatcapfig}
\end{figure}
Its behavior is best understood in terms of the dimensionless parameter $x=T/T_c$,
\beq
C(x,\cN) = \frac{2\cN x \Li_2(\lambda)}{\cW_N}-\frac{\cN x^2}{\lambda \cW_N}
\ln (1-\lambda)\left(\frac{\partial\lambda}{\partial x}\right)_\cN.
\label{heatcap}
\eeq
The specific heat, $C/\cN$, is displayed as a function of $x=T/T_c$ in figure \ref{heatcapfig} 
for $\cN$ ranging from one hundred to one hundred thousand shells. There is no true phase 
transition in this system. At low temperatures, the fugacity in \eqref{fugacity} behaves roughly 
as $1-\cN^{-1}$ and the second term in the expression for the heat capacity in \eqref{heatcap} 
becomes vanishingly small. Thus in the limit as $\beta \rightarrow \infty$, using $\Li_2(1)=
\pi^2/6$ we find
\beq
\frac C\cN  \approx \frac{\pi^2 kT}{3\cN\hbar/l}.
\eeq
(see  \cite{ket96,ing98}). On the other hand, at high temperatures the fugacity behaves as
\beq
\lambda = 1-e^{-\cN\beta\hbar/l}
\eeq
and the specific heat approaches unity in the limit as $\beta\rightarrow 0$.

We now turn to the entropy, which may be computed from
\beq
S = \ln \Xi + \beta U -\mu\beta \cN.
\eeq
Expanding $\ln\Xi$ in a power series and applying the transformation in \eqref{mellinbarnes} 
we arrive at
\beq
\ln \Xi = -\beta M_0 -\ln (1-\lambda)+\frac{\Li_2(\lambda)}{\beta\hbar/l}
\eeq
and combine this expression with \eqref{mass}. Then the entropy function can be given in 
terms of the ratio $x=T/T_c$ as
\beq
S(\beta,\cN) =  \frac{2x\cN}{\cW_N} \Li_2(\lambda) - \ln(1-\lambda) -
\cN \ln \lambda.
\label{entropy}
\eeq
\begin{figure}
\includegraphics[width=7cm]{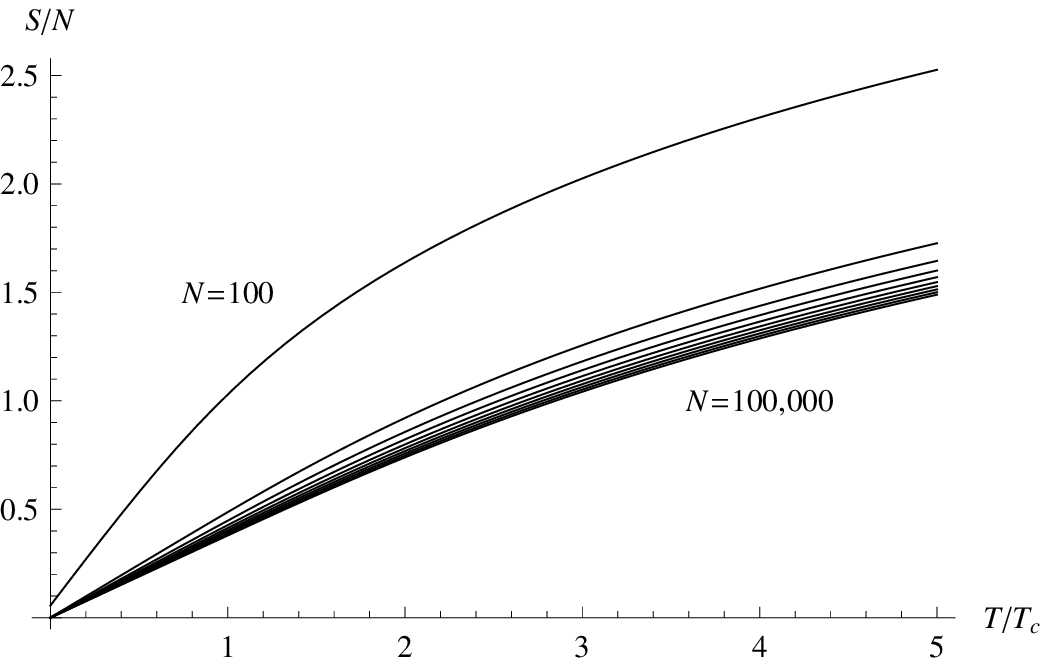}
\caption{$S/\cN$ as a function of $T/T_c$.}
\label{entropyfig}
\end{figure}
The specific entropy is shown in figure \ref{entropyfig} for $\cN$ ranging from one hundred to 
one hundred thousand shells. It is seen to decrease with increasing $\cN$. In the limit of low 
temperatures, {\it i.e.,} as $x\rightarrow 0$, the leading behavior of the entropy is as 
\beq
S \approx \frac{\pi^2}{2\beta\hbar/l} + \ln \cN + 1
\eeq
This is not vanishing, but the specific entropy, $S/\cN$, does vanish in the thermodynamic 
limit as the temperature approaches zero and so it is consistent with a generalized version 
of the third law. If we now consider this function in terms of its natural variables, $M$ and 
$\cN$ we find from \eqref{mass} that in this same limit
\beq
\frac\pi{\beta\hbar/l} \approx \sqrt{\frac{6(M-M_0-\cN\varepsilon_0)}
{\hbar/l}},
\eeq
showing that the black hole temperature vanishes as the black hole mass approaches $M_0 + 
\cN \varepsilon_0$, at which point the entropy approaches $\ln \cN$. In fact, written 
in terms of the black hole mass and the number of shells, the entropy is
\beq
S \approx \frac\pi 3 \sqrt{\frac{6(M-M_0-\cN\varepsilon_0)}{\hbar/l}} 
+ \ln \cN +1.
\eeq
Small $x$ is the regime in which Bose condensation plays a 
dominant role. At temperatures well below the critical temperature, where a large fraction of 
the shells are in the ground state, the first term on the right becomes negligible and the entropy 
is dominated by 
its logarithmic behavior. Thus even though a quantization of Einstein's gravity yields the 
usual black hole results in the thermodynamic limit and at high temperatures, its behavior 
outside this regime may present interesting features peculiar to quantum statistics. In particular, 
we see that BTZ black holes turn cold once condensation sets in. This is the ``small'' black hole 
or short distance regime in 2+1 dimensions. 

\begin{figure}
\includegraphics[width=7cm]{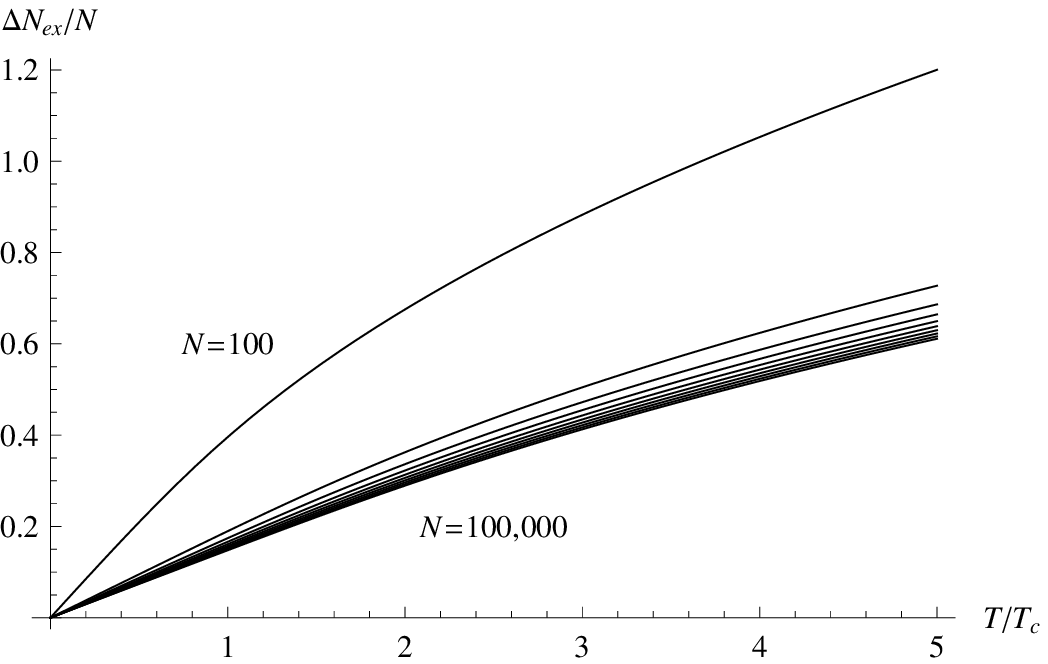}
\caption{$\Delta N_\text{ex}/\cN$ as a function of $T/T_c$.}
\label{fluct}
\end{figure}

Finally, for completeness, we address the mean square fluctuations. We are interested in 
evaluating 
\beq
\Delta N_\text{ex}^2 = -\frac 1\beta\sum_{j=1}^\infty \frac{\partial N_j}
{\partial \varepsilon_j}= \sum_{j=1}^\infty\sum_{r=1}^\infty r \lambda^r 
e^{-\beta \hbar rj/l}.
\eeq
The sums can be evaluated with the help of the representation in \eqref{mellinbarnes} and 
one obtains
\beq
\Delta N_\text{ex}^2  = \frac{\Li_2(\lambda)}{(\beta\hbar/l)^2}.
\eeq
The fluctuations are plotted as a function of $x=T/T_c$ in figure \ref{fluct} for one hundred, 
one thousand and one hundred thousand shells. The uppermost curve represents $\cN$ ranging from 
one hundred to one hundred thousand shells. and the 
fluctuations are seen to decrease with $\cN$. Near the critical temperature, they fall off as 
$1/\ln\cN$.

\section{Discussion}

Although well adapted to problems possessing high symmetry such as in cosmology and 
spherical collapse, the canonical approach suffers from several ambiguities. We have dealt with 
them explicitly before arriving at the results quoted in the introduction, but only within the 
context of the specific class of models with which we work. Thus canonical quantization 
cannot be expected to yield the final theory of quantum gravity as it stands, but its 
results should be taken seriously for models in which the ambiguities can be resolved. 
One advantage of our approach to the quantum properties of black holes is that the fundamental 
degrees of freedom and the black hole microstates are given a transparent meaning. Another is 
that the same picture holds for higher dimensional black holes in Einstein's general relativity 
both with and without a cosmological constant. 

It seems remarkable at first sight that the spherical midisuperspace models of collapse should 
yield the Bekenstein-Hawking entropy of the black hole, considering that all non-spherically 
symmetric states have been eliminated by the symmetry reduction and therefore do not contribute 
to the entropy. We cannot definitively tell why this is so, but we speculate that non-spherically 
symmetric states may not describe a regular horizon (with constant surface gravity) and so do not 
contribute to the entropy of a static black hole. However, this conjecture requires further 
clarification and the question seems worthy of further investigation.

The point we have made in this paper is that the statistics applied in counting black hole microstates,
together with the specific models of the microstates, can play a crucial role in the behavior 
of a black hole at low temperatures and outside the thermodynamic limit. We have demonstrated 
this explicitly in the case of the BTZ black hole for which Bose statistics must be 
applied to correctly recover its thermodynamic properties. We have argued that the BTZ black hole 
condenses in the sense that there is a critical temperature below which a significant fraction 
of the black hole mass is in the ground state. 

Our results can be summarized as follows: (i) the 
condensation temperature grows almost linearly with the number of shells and can be quite large, 
(ii) the specific heat grows linearly below the condensation temperature but slowly above it, 
approaching unity in the limit of very high temperatures and (iii) the Bekenstein-Hawking area 
law is obeyed only in the thermodynamic limit, outside of which the entropy deviates from the 
area law, approaching $\ln \cN$ in the limit as $T\rightarrow 0$, which occurs at finite mass.  
According to the results of 
\cite{jac95}, this would signal a breakdown of general relativity at low enough temperatures. 
This seems to be an interesting direction for future research. 

Of considerable phenomenological importance of course is the question of whether these results 
are reproduced by realistic black holes in higher dimensions and precisely how. If they are, 
then Bose condensed primordial black holes would make natural Dark Matter candidates. We 
will examine the consequences of the Bose condensation described in this paper for gravity at 
short distances in a future publication.
\bigskip\bigskip

\noindent{\bf Acknowledgements}
\bigskip

\noindent LCR Wijewardhana was supported in part by the U.S. Department of Energy Grant No. 
DE-FG02-84ER40153.

\end{document}